\documentclass[twocolumn,aps,prb]{revtex4}
\usepackage{graphics,float}

\def\build#1_#2{\mathrel{\mathop{#1}\limits_{#2}}}
\begin{document}

%\twocolumn[\hsize\textwidth\columnwidth\hsize\csname
%@twocolumnfalse\endcsname

\title{ Effect of nearest neighbor repulsion on the low frequency
phase diagram of a quarter-filled Hubbard-Holstein chain}

\author{Philippe~Maurel} 
\author{Marie-Bernadette~Lepetit}
\email{Marie@irsamc.ups-tlse.fr}

\affiliation{Laboratoire de Physique Quantique \& UMR 5626 du CNRS,
Universit\'e Paul Sabatier, F-31062 Toulouse Cedex 4, France }

\date{\today}

\begin{abstract}
We have studied the influence of nearest-neighbor (NN) repulsions on
the low frequency phases diagram of a quarter-filled Hubbard-Holstein
chain. The NN repulsion term induces the apparition of two new long
range ordered phases (one $4k_F$ CDW for positive $U_{e\!f\!f} =
U-2g^2/\omega$ and one $2k_F$ CDW for negative $U_{e\!f\!f}$) that did
not exist in the $V=0$ phases diagram. These results are put into
perspective with the newly observed charge ordered phases in organic
conductors and an interpretation of their origin in terms of
electron molecular-vibration coupling is suggested.
\end{abstract}
\maketitle

%\vskip2pc]

%==========BODY OF PAPER =========================================

\section{Introduction}
It is well know that low dimensional systems are susceptible to
structural distortions driven by electron-phonons interactions. The
most commonly studied phonons-driven instability is the metal to
insulator Peierls transition in one-dimensional (1D) conductors.  The
insulating state is a periodic modulation of bonds charge density
(BDW) associated to a lattice distortion.  Such instabilities are
driven by the coupling between the electrons and the inter-site
phonons modes, the interaction been essentially supported by a
modulation of the hopping integrals between two nearest neighbor (NN)
sites. The consequence is an alternation of the bond orders while the
on-site charge remains homogeneous on the whole system. 

When a ``site'' stands for a complex system with internal degrees of
freedom, there is another important type of electron-phonons (e-ph)
interaction, namely the one that couples the electrons with the
internal phonons modes of each ``site''.  This is, for instance, the case in
molecular crystals where a site represents a whole molecule.  In such
cases the totally-symmetric (Raman active) molecular vibrational modes
can couple to the system electronic structure. Holstein~\cite{hols}
was one of the first to understand the importance of such e-ph
coupling, showing that it may lead to a self trapping of the
conducting electrons by local molecular deformations. In the 60's
Little~\cite{little} even suggested that intra-molecular vibrations could be
responsible for super-conductivity in the organic
conductors.  More recently it was again proposed to be
the super-conductivity mediator in fullerides-based
systems~\cite{c60}.

Even-though the electron molecular-vibrations (e-mv) has now been
excluded as super-conductivity mediator in organic conductors, a simple
analysis shows that they should in anyway be relevant to these
systems. Indeed, the conductivity-supporting molecules, such as the
$TMTTF$ or $TMTSF$ -based molecules, the $TCNQ$-based molecules, the
$M(dmit)_2$-based molecules, etc, have a certain number of
characteristics in common. They are large, planar, conjugated and
based on organic cycles, all characteristics favorable to a strong
coupling of the conduction electrons (which, belonging to the $\pi$
conjugated system of the molecules, are strongly delocalized on the
molecular skeleton) with the molecular totally symmetric ($A_g$)
vibrational modes. This analysis is fully supported by Raman
spectroscopy measurements~\cite{vib,review_1D} which assert both the
existence of low frequency vibrational modes (associated with ring
angular deformations) and e-mv coupling constants belonging to the
intermediate regime.

It is widely admitted that the simplest pertinent model for describing
the 1D organic conductors electronic structure is the extended Hubbard
(eH) model with NN bi-electronic repulsions. The present study seeks
therefore at studying the combined effects of the electron correlation
within the eH model and the e-mv interactions, in a quarter-filled 1D
chain, relevant for organic 1D conductors. The e-mv problem have been
largely addressed in the case of the one-dimensional half-filled
chain~\cite{hf}. Quarter-filled systems have been treated in several
regimes. In the weak coupling regime renormalization group (RG)
approaches~\cite{rg1} show that the transition line between the
Luttinger Liquid~\cite{ll1,rev_ll} (LL) phase and the 
Luther-Emery~\cite{le,rev_ll} (LE) phase (gaped spin channel and
dominating $2k_F$ charge fluctuations) is displaced toward positive
$g_1$ parameters when the e-mv coupling increases. In addition, the
Luttinger Liquid parameters are renormalized by the e-mv interactions.
In the adiabatic and small inter-site repulsion
regime~\cite{adiab1,adiab2}, small systems diagonalizations exhibit
three different phases, one uniform phase at small e-mv coupling,
associated with LL, one $2 k_F$ charge density wave (CDW) phase for
large enough e-mv coupling and small values of the on-site electron
correlation, and one $4k_F$ CDW phase for large enough e-mv coupling
and on-site repulsion. When inter-site repulsion is omitted, we have
studied~\cite{ph1} the whole phases diagram as a function of both the
phonons frequency, the e-mv coupling and the on-site correlation
strength. We have shown that the dependence of the phases diagram to
the phonons frequency is crucial. Indeed, while for high frequencies
(corresponding to the highest $A_g$ molecular vibrations of the
Bechgaard salts) the phases diagram is very poor and well reproduced by
the weak coupling approximation, for low phonons frequencies
(corresponding to the lowest $A_g$ molecular vibrations of the
Bechgaard salts), the phases diagram is on the contrary very rich. At
small e-mv coupling, and in agreement with the RG results, we found LL
and LE phases with renormalized parameters. In the intermediate
coupling regime, we found at surprisingly small values of the on-site
repulsion (from $U/t\sim 2$), a metallic phase with dominating $4k_F$
CDW fluctuations. For large e-mv coupling we found polaronic phases
where the electrons are self-trapped by the molecular deformations,
either by pairs (low electron-correlation regime) or alone (large
electron-correlation regime).

The NN bi-electronic repulsions are crucial for a reliable description
of the 1D organic conductors.  The present paper will therefore study
the interplay between the e-mv and the NN bi-electronic repulsion
within an extended Hubbard model. In regards to the previous results
we will limit us to the low phonons frequencies where we can expect
significant effects to occur.

The next section will be devoted to the model description and the
computational details. Section 3 will present the results and section
4 will discuss their relevance to the organic conductor physics. The
last section will conclude.

\section{Computational details and model}

\subsection{Model}
The simplest way to couple dynamically dispersion-less molecular
vibrations to the electronic structure is through local harmonic
oscillators and linear e-mv coupling. We will therefore use an
extended-Hubbard-Holstein model (eHH). If $U$ stands for the on-site
repulsion, $V$ for the nearest-neighbor coulomb repulsion and $g$ for
the e-mv coupling constant the eHH model can be written as $H_{e} +
H_{ph} + H_{e-mv}$ with
\begin{eqnarray*}  %\hspace*{-3eM}
H_{e} &=& t\sum_{i,\sigma}{(c_{i+1,\sigma}^{\dagger}c_{i,\sigma}}+
h.c.)  +
U\sum_{i}{n_{i,\uparrow}n_{i,\downarrow}}+V\sum_{i}{n_{i}n_{i+1}}\\
H_{ph} &=&\omega\sum_{i}{(b_{i}^{\dagger}b_{i}+1/2)}\\ \hspace*{-2eM}
H_{e-mv}&=&g\sum_{i}{n_{i}(b_{i}^{\dagger}+b_{i})}
\end{eqnarray*}
$c_{i,\sigma}^{\dagger }$, $c_{i,\sigma}$ and $n_{i,\sigma}$ are the
usual creation, annihilation and number operators for an electron of
spin $\sigma$ located on site $i$
($n_i=n_{i,\uparrow}+n_{i,\downarrow}$).  $b_{i}^{\dagger}$ and
$b_{i}$ are the intra-molecular phonons creation and annihilation
operators and $\omega$ the phonons frequency.  The energy scale is
fixed by $t=1$.

As noticed in ref.~\cite{ph1}, the on-site part of the Hamiltonian can
be rewritten (apart from constant terms) as
\begin{eqnarray} \hspace*{-5eM} 
\label{hi}&&
\omega \left[ \left( b_{i}^{\dagger} + n_i{g\over \omega} \right)
      \left( b^{\hfill}_{i} + n_i{g\over \omega} \right) \right] - n_i
      {g^2 \over \omega} + \left(U- 2{g^2 \over \omega} \right)
      n_{i,\uparrow}n_{i,\downarrow}
\end{eqnarray}

On may highlight the following points. \begin{itemize}
\item The on-site bi-electronic repulsion term is renormalized by the
e-mv coupling and the effective interaction $U_{e\!f\!f} = U- 2 g^2/
\omega$ becomes attractive in the strong coupling regime.
\item One sees from eq.~\ref{hi} that the phonons and e-ph parts of
the Hamiltonian can be rewritten as a displaced harmonic
oscillator. The noticeable point is that the displacements are
proportional to the sites charge, simulating this way the relaxation
of the molecular geometry as a function of the ionicity of the site.
\item The natural basis for the phonons states is therefore the
eigenstates of the displaced oscillators. Such a vibronic
representation is not only very physical, but also particularly suited
for the representation of the low energy physics.  Despite the fact
that one would have a complete basis set for each value of the sites
occupation, the necessity to work in a truncated representation lift
the problem of over-completeness.
\item One should also notice that there is a strong renormalization of
the hopping integrals between the initial and final vibronic states of
two neighboring sites, by the Franck-Condon factors.  In this
representation the Franck-Condon factors correspond to the overlaps
between the vibronic states associated with $\pm 1$ occupation
numbers. As physically expected, when an electron hopes between two
sites vibronic ground states, the hopping integral is exponentially
renormalized by the displacement~: $t \longrightarrow t \;
\exp{\left[-\left(g/\omega\right)^2\right]}$.  The direct consequence
is an increased tendency to electron localization. Indeed, vibronic
high energy states need to be summon up for delocalization processes
when the e-mv coupling is large.
\item The pertinent e-mv coupling parameter is not $g$ but rather
$g/\omega$, in the light of which it becomes clear that only
vibrations of low frequencies may produce significant effects other
than a simple renormalization of the pure electronic interactions. The
model pertinent parameters are therefore $U/t$, $V/t$, $\omega/t$ and
$g/\omega$
\end{itemize}

\subsection{Computational details} \label{ss:cd}
The calculations have been carried out using the infinite system
density-matrix renormalization group (DMRG) method~\cite{dmrg} with
open boundary conditions. 

\begin{figure}[h]
\centerline{\resizebox{6cm}{4cm}{\includegraphics{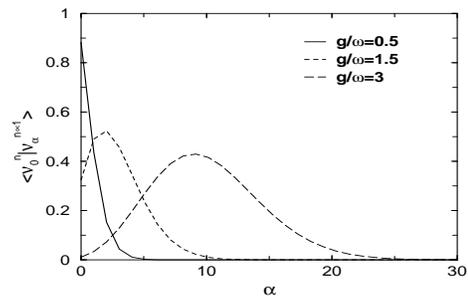}}}
\vspace{5mm}
\caption{Overlap between the vibronic ground state $\nu_0^n$ 
of a site supporting $n$ electrons and the $\alpha^{\rm th}$ vibronic state 
$\nu_{\alpha}^{n\pm 1}$of a site supporting $n\pm 1$ electrons  .}
\label{fig:rec} \vspace*{2eM}
\end{figure}

%One of the consequences of working in a
%truncated vibronic basis set is that it destroys any further possibility 
%to separate the electronic degrees of freedom from the vibrational ones.
Since an infinite number of phononic quantum states lives on each site
we have truncated the phonons basis set according to the previous
section analysis, that is we kept only the two lowest vibronic states
on each site (i.e. the two lowest eigenstates of the on-site
Hamiltonian).  This choice is physically reasonable since (i) we work
at $T=0$ and therefore only the lowest vibronic states are expected to
be involved, (ii) the molecules form well defined entities that are
only perturbatively modified by the presence of their neighbors.  As
already mentioned, when an electron hopes between two nearest
neighbors sites the hopping integral is renormalized by the
Franck-Condon factors, i.e. the overlap between the initial and final
vibronic states of the sites, $\langle (n, \nu, Sz)_i~; (n^\prime,
\nu^\prime, Sz^\prime)_{i+1}|H| (n-1, \mu, Sz\pm 1/2)_i~; (n^\prime+1,
\mu^\prime, Sz\mp 1/2)_{i+1}\rangle = t \, \langle \nu|\mu\rangle \,
\langle \nu^\prime|\mu^\prime\rangle$ ($n$ and $n^\prime$ being the
number of electrons on sites $i$ and $i+1$ respectively, $\nu$ and
$\mu$, $\nu^\prime$ and $\mu^\prime$ the phonons states and $Sz$,
$Sz^\prime$ the spin projection quantum numbers). Figure~\ref{fig:rec}
shows the overlap between the vibronic ground state of a site
supporting $n$ electrons and the vibronic states of the same site
supporting $n\pm 1$ electrons.  As can be seen, when $g/\omega$ is
small the overlap, and therefore the Franck-Condon factor, decreases
very quickly with the number of bosons, thus only the first few
vibronic states are involved and the truncation is totally
pertinent. This fact is confirmed by exact calculations on small
systems (4 sites) where four phononic quantum numbers have been
considered. For instance, the weight of the 3 and 4 bosons
contributions in the wave functions is only $0.012$ for $g/\omega =
0.5$, $U/t=4$ and $U/V=4$. When $g/\omega$ increases, the maximum of
the Franck-Condon factors is rapidly displaced toward very large
number of bosons. These vibronic states being strongly hindered by
their large vibrational energy, they have a small weight in the wave
function and the system tends to localize. For instance for $g/\omega
= 3$, $U/t=4$ and $U/V=4$, the weight of the 3 and 4 bosons
contributions in the 4 sites system is only $1.2\times 10^{-3}$. In
the intermediate region the contribution of intermediately large
phonons states (with occupations $3,4,5$) is not as negligible ($0.3$
for $g/\omega = 1.5$, $U/t=4$ and $U/V=4$), and the basis set
truncation lowers the total hoping between nearest neighbors sites
and thus increases the system localization. One can therefore expect
that in the intermediate regime the true phases transitions will be
displaced ---~compared to our results~--- toward larger values of the
electron-phonon interaction. It is however clear that any truncated
basis set will have a great deal of problem to accurately treat
systems too close to a phases transition since the effective softening
of the frequency near the transition translates into the implication
of a quasi-infinite number of phonons states. From the above analysis,
one can be quite confident in the quality of the numerical results and
in particular in the different phases found in this work, provided
that the exact position of the transitions is not seek at, in the
intermediate coupling region. In order to estimate more precisely the
transition displacement due to the basis set truncation, we have run
additional calculations using up to three bosons states per site
occupation number (that is 12 on-site states) for a set of chosen
electronic parameters ($U/t=4, \quad U/V=4$) and all values of the
electron-phonon coupling constant.

In order to characterize the
phases diagram, we have computed the charge and spin gaps, defined as
usual~:
$$\Delta_{\rho} = E_{0}(2N,N+1,0)+E_{0}(2N,N-1,0)-2E_{0}(2N,N,0)$$ and
$$\Delta_{\sigma}=E_{0}(2N,N,1)-E_{0}(2N,N,0)$$ where
$E_{0}(N_s,N_e,Sz)$ is the ground state (GS) energy of a system of
$N_e$ electrons, $N_s$ sites and spin projection $Sz$.  In addition,
we have computed the charge-charge, spin-spin correlation functions~:
$c_A(j)= \langle \left(A_{0}-\langle A_{0}\rangle\right)
\left(A{j}-\langle A_{j}\rangle\right) \rangle$ where $A$ stands
either for the number operator, $n$, or for the spin projection
operator, $Sz$, and the on-site singlet correlation function~:
$c_{Sg}(j)=\langle {Sg_{0}}^{\dagger} Sg_{j}\rangle$ where
${Sg_i}^{\dagger}= c_{i,\uparrow}^{\dagger}c_{i,\downarrow}^{\dagger}$
and $Sg_i$ are the singlet creation and annihilation operators on a
site.

The properties calculation have been computed using $255$ states per
renormalized block, whereas for the gaps calculations we have used a
double extrapolation (i) on the systems size, and (ii) on the number,
$m$, of states kept, using $m=100,150$ and $255$.  In all calculations
we have computed systems of size up to 80 sites and extrapolated to
the infinite chain.

To have more information on the localized phases wave functions we
have also computed the density matrices at the central sites and
performed exact diagonalization of small systems.

\section{Result}
The present work explores the whole range of the on-site repulsion
strength and of the e-mv coupling. The vibration frequency have been
chosen to be $\omega/t=0.2$ for the reasons already exposed in the
preceding paragraphs. The phases diagrams have been computed for two
values of the nearest-neighbor versus on-site repulsions ratio $V/U$
which are recognized to be generic for the 1D organic
conductors~\cite{corr_v}, namely $V=U/4$ and $V=U/2$.

\begin{figure}[h]
\centerline{\resizebox{8cm}{6cm}{\includegraphics{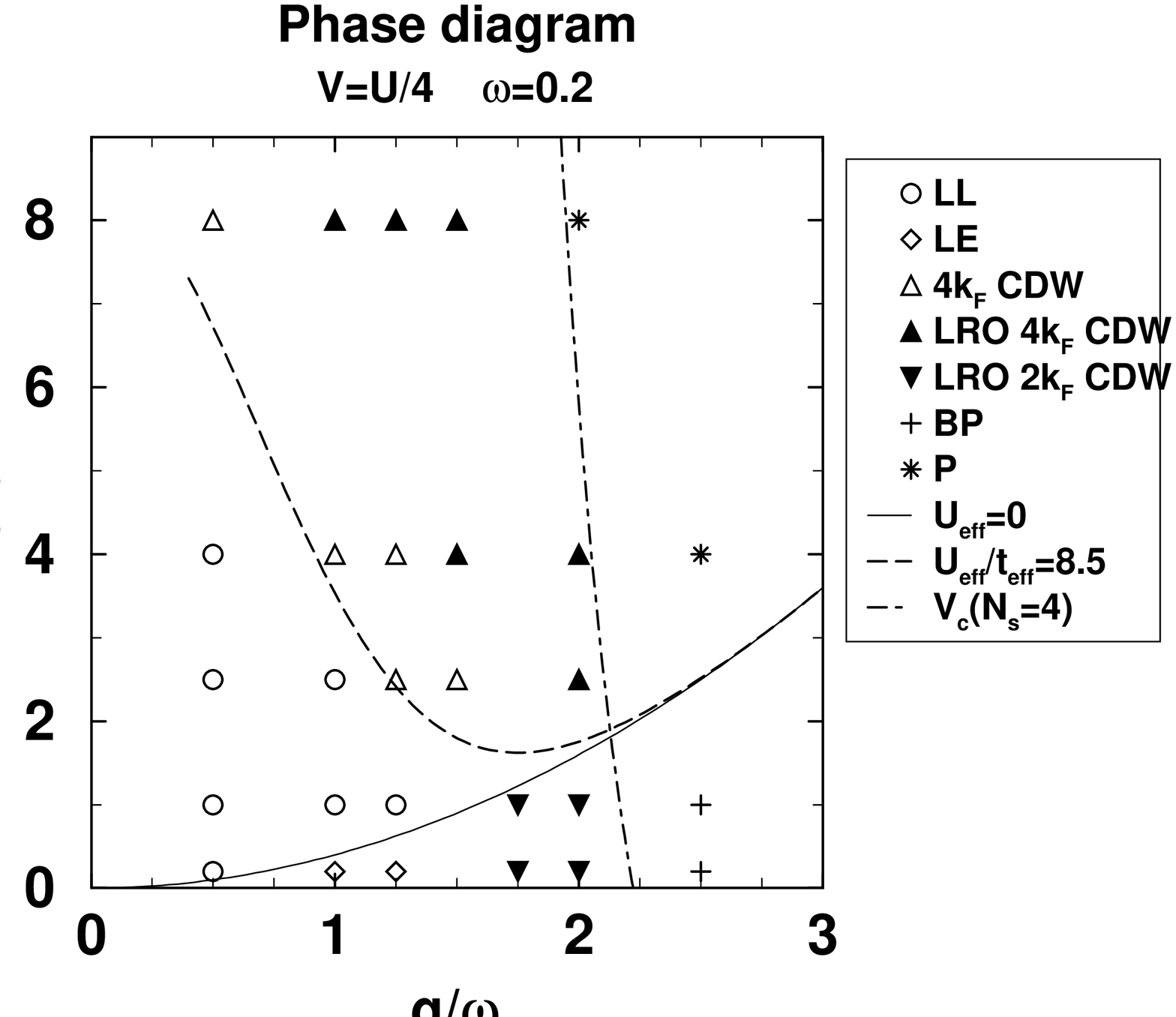}}}
\vspace{5mm}
\caption{Phases diagram of the extended Hubbard-Holstein model for
$V=U/4$ and $\omega/t=0.2$.  }
\label{fig:diag1} \vspace*{2eM}

\centerline{\resizebox{8cm}{6cm}{\includegraphics{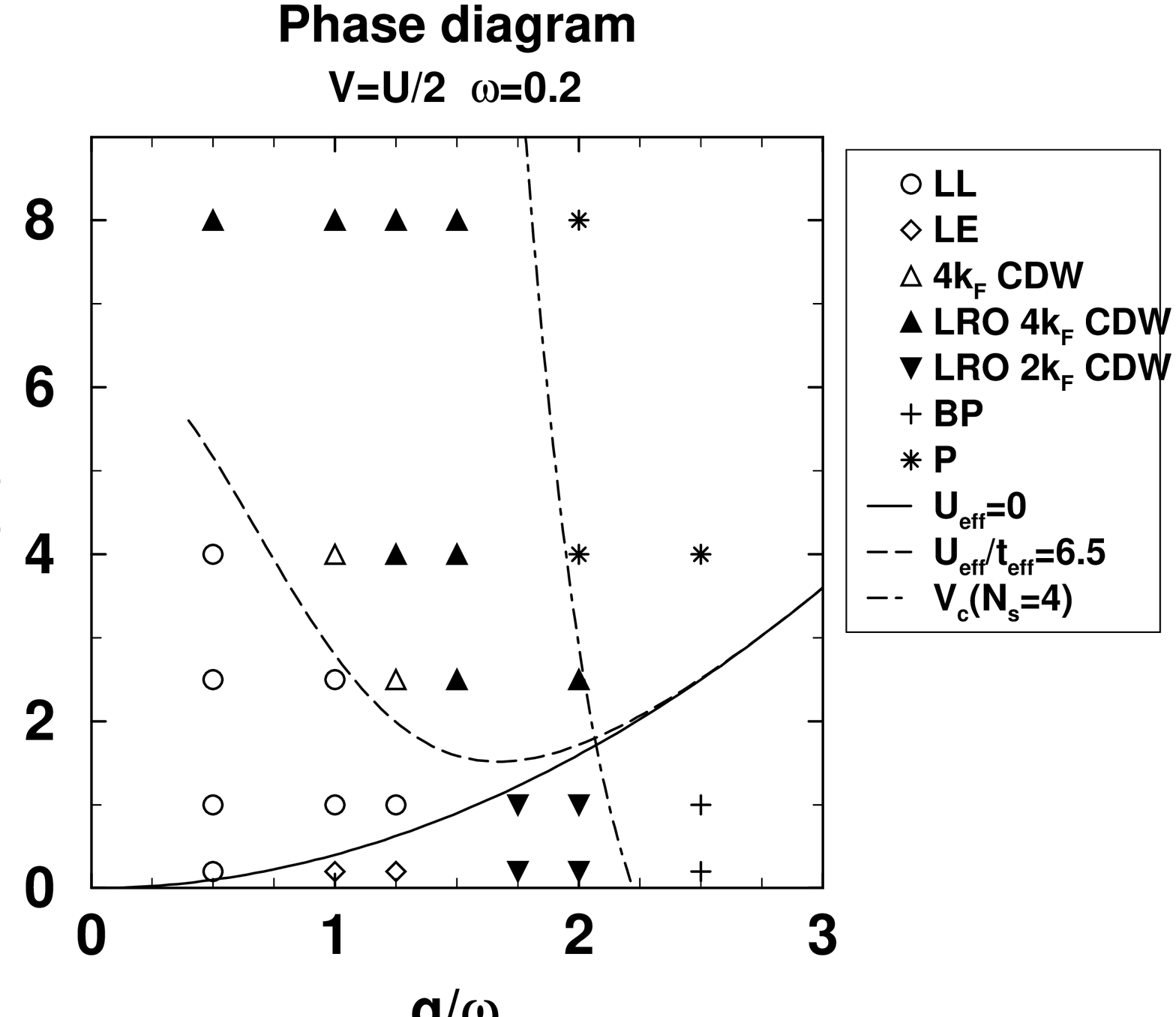}}}
\vspace{5mm}
\caption{Phases diagram of the extended Hubbard-Holstein model for
$V=U/2$ and $\omega/t=0.2$.  }
\label{fig:diag2}
\end{figure}

Figure~\ref{fig:diag1} and~\ref{fig:diag2} report the phases diagrams
for $V=U/4$ and $V=U/2$ as a function of $g/\omega$ and $U/t$.  The
two diagrams present the same general features with seven different
phases. The major effect of the introduction of NN repulsion in the
Hubbard-Holstein model is to stabilize two new phases in the e-mv
intermediate coupling regime. From another point of view, the
inclusion of the e-mv coupling in the extended Hubbard model has
similar consequences as its inclusion in the pure Hubbard model; that
is~: the apparition of polaronic and bi-polaronic phases in the strong
coupling regime, the apparition of a $4k_F$ CDW phase in the
intermediate regime for extremely low values of the on-site repulsion.

To summarize, in the weak coupling regime one find both the Luttinger
Liquid phase for $U_{e\!f\!f}>0$ and the Luther Emery phase for
$U_{e\!f\!f}<0$. In the strong coupling regime one has the polaronic
($U_{e\!f\!f}>0$) and bi-polaronic ($U_{e\!f\!f}<0$) phases where the
electrons are self-trapped (alone or by pairs) by the molecular
geometry deformations. In between these two regimes, that is for
intermediate e-mv coupling, one finds the two new phases. The first
one is an insulating long-range ordered $4k_F$ CDW phase which
develops at the extend of the metallic $4k_F$ CDW phase for
$U_{e\!f\!f}>0$. The second one is an insulating long-range ordered
$2k_F$ CDW phase which develops for $U_{e\!f\!f}<0$ at the expends of
the localized bi-polaronic phase.  It is noticeable the
$U_{e\!f\!f}=0$ line seems to remain a strict phases boundary. On the
contrary the other frontiers have been shifted. For $U_{e\!f\!f}>0$
the localized phases are enhanced and the delocalized ones reduced. On
the contrary, for $U_{e\!f\!f}<0$ the bi-polaronic phase is reduced.

\subsection{The $U_{e\!f\!f}>0$ phases}

\subsubsection{Luttinger liquid phase}
For small values of $g/\omega$, up to intermediate ones if the on-site
repulsion $U$ is not too large, one finds, both in the $U/V=4$ and
$U/V=2$ cases, the expected LL phase.  The computed charge and spin
correlation functions exhibit power law behavior with dominant $2k_F$
SDW fluctuations and sub-dominant CDW fluctuations.  The spin and
charge gaps extrapolate nicely to zero to numerical accuracy.

Similarly to what happens in the pure extended Hubbard
model~\cite{eH}, the $2k_F$ SDW fluctuations and the $4k_F$ CDW
fluctuations are enhanced by increasing values of the NN repulsions.
\begin{figure}[h]
\centerline{\resizebox{6cm}{6cm}{\includegraphics{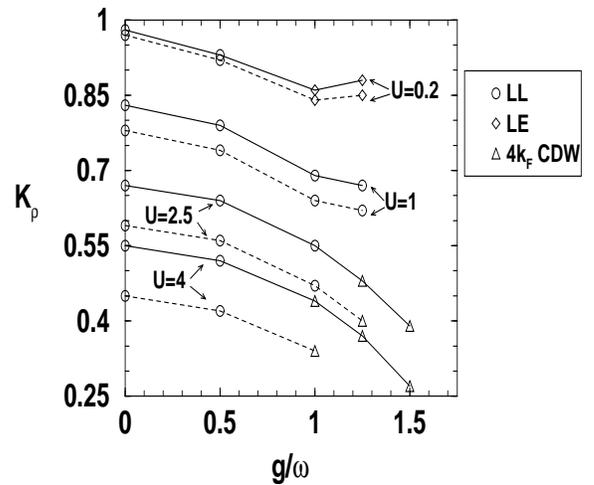}}}
\vspace{5mm}
\caption{ $K_{\rho}$ as a function of $g/\omega$ for $U=0.2,1,2.5,4$,
$V=U/4$ (solid lines) and $V=U/2$ (dashed lines). The symbols follow
the same pattern as in the phases diagrams.}
\label{fig:kp}
\end{figure}
From the charge structure factor $S_\rho(q)$ we have computed the LL
$K_\rho$ parameter as
\begin{eqnarray*}
K_{\rho} &=& \pi \left.{d\over dq} S_{n}(q) \right|_{q=0}
\end{eqnarray*}
Figure~\ref{fig:kp} reports the $K_\rho$ parameter as a function of
both $U/V$ and $g/\omega$. Once again the results are a simple
superposition of the $K_\rho$ reduction effect due to the NN
repulsions and the reduction effect due to the e-mv coupling.

\subsubsection{The $4k_F$ CDW phase}
The LL phase is bordered, both for $V=U/4$ and for $V=U/2$, by a
metallic phase presenting dominating $4k_F$ charge fluctuations. This
phase have very similar characteristics as the $4k_F$ CDW phase found
for $V=0$~\cite{ph1}, that is no charge neither spin gap (to numerical
accuracy), power law decreasing of the charge and spin correlation
functions, dominant $4k_F$ CDW fluctuations and very small values of
$K_\rho$ compared to the purely electronic model. One has for
instance, for $U/t=4$ and $V/t=1$, $K_\rho= 0.28$ when $g/\omega=1.5$
instead of $K_\rho = 0.55$ in the pure electronic eH model (see
fig.~\ref{fig:kp}).  One should however notice that while the $K_\rho$
values remain always larger than the $1/4$ minimal value predicted by
the LL theory~\cite{rev_ll} for metallic behavior, it can be as large
as $0.48$ for $U=2.5$, $g/\omega=1.25$ and $V=U/4$, that is much above
the $1/3$ limiting value predicted by the LL theory for dominant
$4k_F$ CDW fluctuations~\cite{rev_ll}.  Despite the absence of long
range coulombic repulsions, this phase is in fact, in many ways very
alike a Wigner crystal.

Figures~\ref{fig:diag1} and~\ref{fig:diag2} show that the NN
repulsions have a strongly destructive effect on this phase.  Indeed,
it reduces strongly its domain of existence for increasing $g/\omega$.
Compared to the $V=0$ case a new insulating, long-range order (LRO)
$4k_F$ CDW phase has taken a large part of the $g/\omega$ parameters
range of the metallic $4k_F$ CDW phase.  For increasing $V/U$ the
metal to insulator phases transition (MIT) is repelled to smaller
values of $g/\omega$, squeezing the metallic $4k_F$ CDW phase toward
the LL one.

\subsubsection{The LRO $4k_F$ CDW phase}
As $g/\omega$ increases, the system undergoes a metal to insulator
phases transition and the $4k_{F}$ CDW fluctuation phase condensates
into a long range ordered $4k_{F}$ CDW phase.

In order to characterize this new phase, we have computed the
 staggered charge correlation functions $(-1)^{j}{\cal C}_{n}(j)$
 where
 $${\cal C}_{n}(j)=\langle(n_{i} - \bar n)(n_{i+j} - \bar n)\rangle$$
and $\bar n = N_{e}/N_{s} = 1/2$ is the average charge per site. The
associated order parameter is therefore
$$ X_{4k_F}=\lim\limits_{N_{s}\rightarrow+\infty}
\sum_{j}{(-1)^{j}{\cal C}_{n}(j)}$$ In this gaped regime, one should
be careful and clearly distinguish between the correlation functions
${\cal C}_{n}(j)$ and the correlation functions of the observable
fluctuations ${c_{n}}(j)$.  Indeed, while in delocalized phases the
two do not differ, in gaped phases the correlation functions tends
toward a non zero constant as the inter-site distance increases, while
the fluctuations decrease quickly to zero at infinite inter-site
distances~\cite{rev_ll}.

\begin{figure}[htp]
\centerline{\resizebox{6cm}{6cm}{\includegraphics{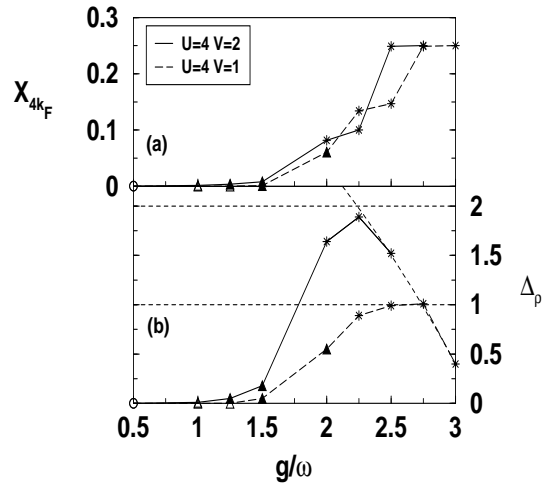}}}
\vspace{10mm}
\caption{(a) Order parameter $X_{4k_F}$ and (b) charge gap as a
function of $g/\omega$ for $U=4$. The solid lines correspond to
$U/V=2$ and the long dashed lines to $U/V=4$. The symbols follow the
same pattern as in the phases diagrams. The short light dashed lines
correspond respectively to $\Delta_\rho = V$ (that is $V=2$ and $V=1$)
and $\Delta_\rho = U_{e\!f\!f}$.}
\label{fig:xnch}
\end{figure}
 
Figure~\ref{fig:xnch} reports both the charge gap $\Delta_\rho$ and
the order parameter $X_{4k_F}$ as a function of $g/\omega$, for the
two values of the NN repulsion.  One sees immediately that the opening
of the charge gap from the metallic $4k_F$ CDW phase is simultaneous
with the formation of the long range order. Similarly the computed
charge fluctuations correlation functions ${c_{n}}(j)$ go from a power
law behavior as a function of increasing inter-site distances to an
exponential behavior. One should note that the spin channel remains
ungaped and the corresponding fluctuations correlation functions
${c_{Sz}}(j)$ decrease as a power law with increasing inter-site
distances.

At the metal to insulator phases transition, that is for intermediate
values of $g/\omega$ ($\simeq 1$), one observes a smooth opening
(exponential like) of both the charge gap and the order parameter~;
opening that seems consistent with a Kosterlitz-Thouless
transition. For large value of $g/\omega$, the order parameter
saturates and the system undergoes a self-trapping transition of the
electrons toward a polaronic phase. It is noticeable that, while the
MIT is very soft, the self-trapping transition is on the contrary
rather sharp.

\subsubsection{The small polaronic phase}
\begin{figure}[h]
\centerline{\resizebox{6cm}{5cm} {\includegraphics{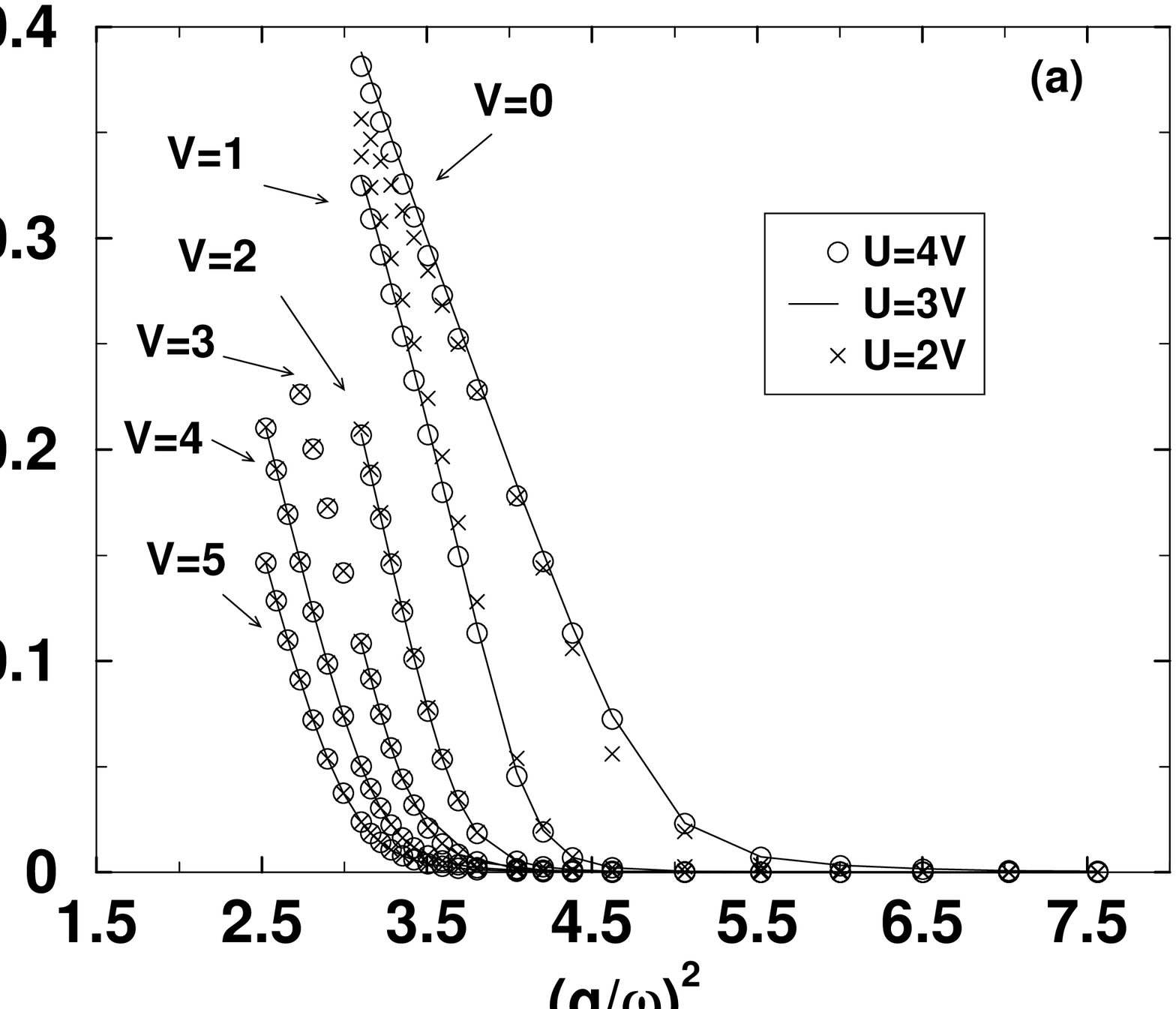}}}
\vspace{5mm}
\hspace*{5mm}\resizebox{6cm}{3.5cm} {\includegraphics{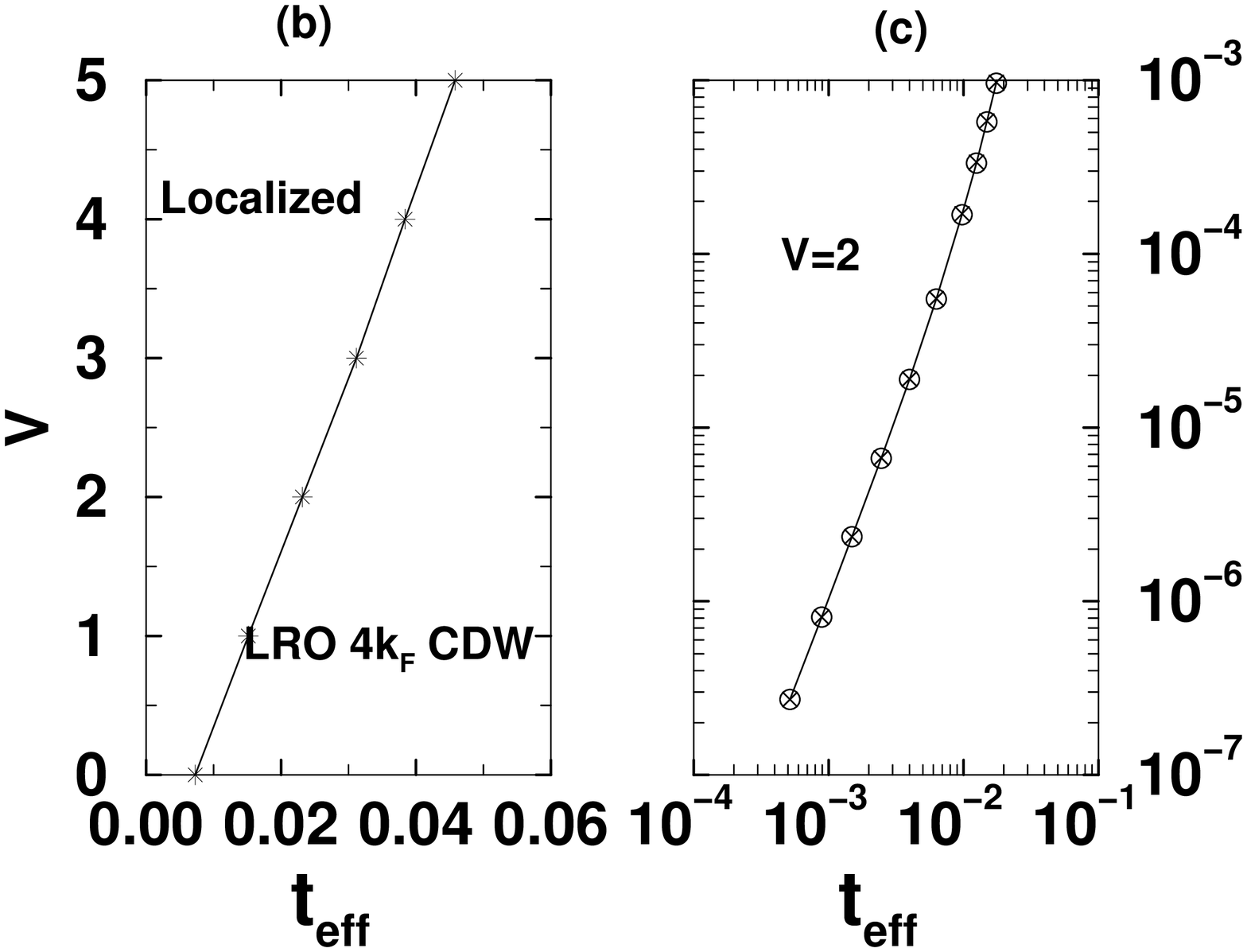}}
\vspace{10mm}
\caption{(a) First excitation energy, $\Delta_{10}$, as a function of
$g^2/\omega^2$ for $4$ sites systems, several value of $V$ and for
$U=4V$ (circles), $U=3V$ (solid lines) and $U=2V$ (crosses). For the $V=0$
case we have chosen $U=8$, $U=4$ and $U=2$. (b) Phase boundary between
the $4k_F$ LRO CDW phase and the polaronic phase as a function of $V$
and $t_{e\!f\!f}$. (c) $\Delta_{10}$ as a function of $t_{e\!f\!f}$ in
a log-log curve for small values of $t_{e\!f\!f}$ (for clarity, we
only show the curves for $V=2$).}
\label{fig:tr_lc_dlc}
\end{figure}

As expected from previous works~\cite{hols,pol2,pol3,ph1}, in the
strong coupling regime but still positive effective on-site repulsion,
$U_{eff}=U-2g^2/\omega$, the system undergoes a transition toward a
polaronic phase where the electrons are self-trapped by the molecular
distortions. This trapping is mediated by the Franck Condon factors,
that strongly renormalize the hopping integrals between low energy
vibronic states. One can remember that the hopping between the ground
vibronic states is renormalized as $t\rightarrow t_{eff}\sim t\
exp(-(g/\omega)^2)$. The ground state of the system is therefore
dominated by configurations such as~:
$$..10101010..$$ where the $1$ stand for sites supporting one electron
and the $0$ stand for empty sites.  The validity of this picture has
been checked both on the GS wave function of small systems (4 and 8
sites, PBC, exact diagonalization) and the on-site density matrix in
the DMRG calculations. On small systems the computed weight of those
configurations is always larger $0.85$ with, for instance, $0.900$ for
$U=4$, $V=2$, $g/\omega=2$ and $4$ sites. On large systems, we have
computed the central sites density matrices and found that the
probability of having double occupations is extremely small, with for
instance, $\rho(\uparrow\downarrow) \le 10^{-9}$ for $U=4$, $g/\omega
= 2.5$ and all values of $U/V$. Coherently, the GS energy per site is
nearly independent of the system size and verify ---~up to at least 4
significant numbers~--- on all the  computed points, the formula
$-N_e/N_s \, g^2/\omega + \omega/2$. Such a GS is strongly
quasi-degenerated due to the equivalence between the odd and even
sites and the different spin configurations. The small splitting is
due to the residual delocalization and therefore scales as
$t_{e\!f\!f}=t\exp(-(g/\omega)^2)$ (see
figure~\ref{fig:tr_lc_dlc}). The spin channel remains ungaped. The
main difference between the present phase and the phase found in the
HH model stays in the charge channel. Indeed, the NN repulsion is
responsible for the opening of a strong gap (see
figure~\ref{fig:xnch}) that did not exist in the $V=0$ case. In fact,
the charge gap scales as the cost to add an extra electron to the
system. In the case of open systems the end sites being always
occupied, $\Delta_\rho$ scales as $\min{\left(U_{e\!f\!f},V\right)}$
(according to whether the extra electron is located on an ``empty''
site or on an already ``occupied'' one) while in periodic systems is
scales as $\min{\left(U_{e\!f\!f},2V\right)}$. The change of behavior
in $\Delta_\rho$ can be clearly seen on figure~\ref{fig:xnch}, where,
for instance, the gap for $U=4$, $U/V=4$ undergoes a saturation to
$\Delta_\rho=V=1$ at the self-trapping transition (that occurs between
$g/\omega =2$ and $g/\omega =2.2$) and then a strong decrease for
$g/\omega\ge \sqrt{15/2}\simeq 2.74$ ($U_{e\!f\!f}=V$), where is
behaves as $U-2g^2/\omega$. One should notice that the full saturation
of the order parameter occurs only after the second transition.

In order to better study the position of the phases transition between
the $4k_F$ LRO CDW and the polaronic phase, we performed exact
diagonalizations on periodic small systems (4 sites, 2 electrons).
One should remember that while the $4k_F$ LRO CDW phase has a
non-degenerated GS, the polaronic phase presents a quasi-degenerated
GS, the degeneracy lifting being of the order of magnitude of
$t_{e\!f\!f}$. We have therefore used the first excitation energy,
$\Delta_{10}$, as a criteria for the phases transition.
Figure~\ref{fig:tr_lc_dlc}(a) reports $\Delta_{10}$ as a function of
$g^2/\omega^2$ for different values of $U$ and $V$. One first notice
that the excitation energies depend in a negligible way on on-site
repulsions. On the contrary it does strongly depend on the NN
repulsion. Going from very large e-mv coupling to smaller values, the
excitation energy first increases as a power law of
$t_{e\!f\!f}=t\exp(-g^2/\omega^2)$ (see figure~\ref{fig:tr_lc_dlc}(c))
in the polaronic phase then linearly as a function of $g^2/\omega^2$
in the $4k_F$ CDW LRO phase. Decreasing $g^2/\omega^2$ to even lower
values this excitation energy should go through a maximum and then
decrease back to zero at the MIT. The location of the phases 
transition between the $4k_F$ LRO CDW and the polaronic phase have
been evaluated as the point where the linearly extrapolated excitation
energy of the $4k_F$ LRO CDW phase crosses the zero axis.
Figure~\ref{fig:tr_lc_dlc}(b) reports the phase boundary as a function
of $t_{e\!f\!f}$ and $V$. One sees immediately that it follows a
perfectly linear curve that can be fitted as $V=129.36t_{e\!f\!f} -
0.93t$, in these coordinates. This curve have been reported on the
phases diagrams (figures~\ref{fig:diag1} and~\ref{fig:diag2}) as the
$V_c$ curves. One sees immediately that the phases transition position
is very weakly dependent of the system size (as expected from such
localized systems) and that the small systems estimations work pretty
well for the infinite systems.

\subsection{The $U_{e\!f\!f}<0$ phases}

\subsubsection{The Luther-Emery phase}
For negative values of $U_{e\!f\!f}$ and small values of $g/\omega$,
we found a metallic phase for which all spin, charge and on-site
singlet fluctuations correlation functions decrease with the
inter-site distances, as power law. All three correlation functions
exhibit dominating $2k_{F}$ components. In all computed cases, the
$2k_{F}$ CDW fluctuations have the largest amplitudes.  The main
effect of the NN repulsion is to increase the amplitude of the CDW
fluctuations and to strongly decrease the amplitude of the on-site
singlet fluctuations.  The charge gaps clearly extrapolate to zero,
whereas we found a very small gap in the spin channel
($\Delta_{\sigma}\sim 0.002-0.003$).  This fact is not incompatible
with the behavior of the spin-spin correlation function, since very
small gaps means very large correlation lengths of the order of
magnitude of $\Delta_{\sigma}^{-1}$. The expected exponential behavior
of the spin correlation functions should therefore take place at
inter-site distances larger than the computed chain lengths. One can
easily recognize in this phase a weakly gaped Luther-Emery phase.  The
values of $K_{\rho}$ extracted from the charge structure factors are
strongly reduced compare to the values of the purely electronic model,
in a similar way as what has been found in the $V=0$ case. It should
be noted that $K_{\rho}$ always remains lower than $1$ in agreement
with dominant CDW fluctuations.

\subsubsection{The LRO $2k_{F}$ phase}
\begin{figure}[h]
\centerline{\resizebox{5cm}{4cm}{\includegraphics{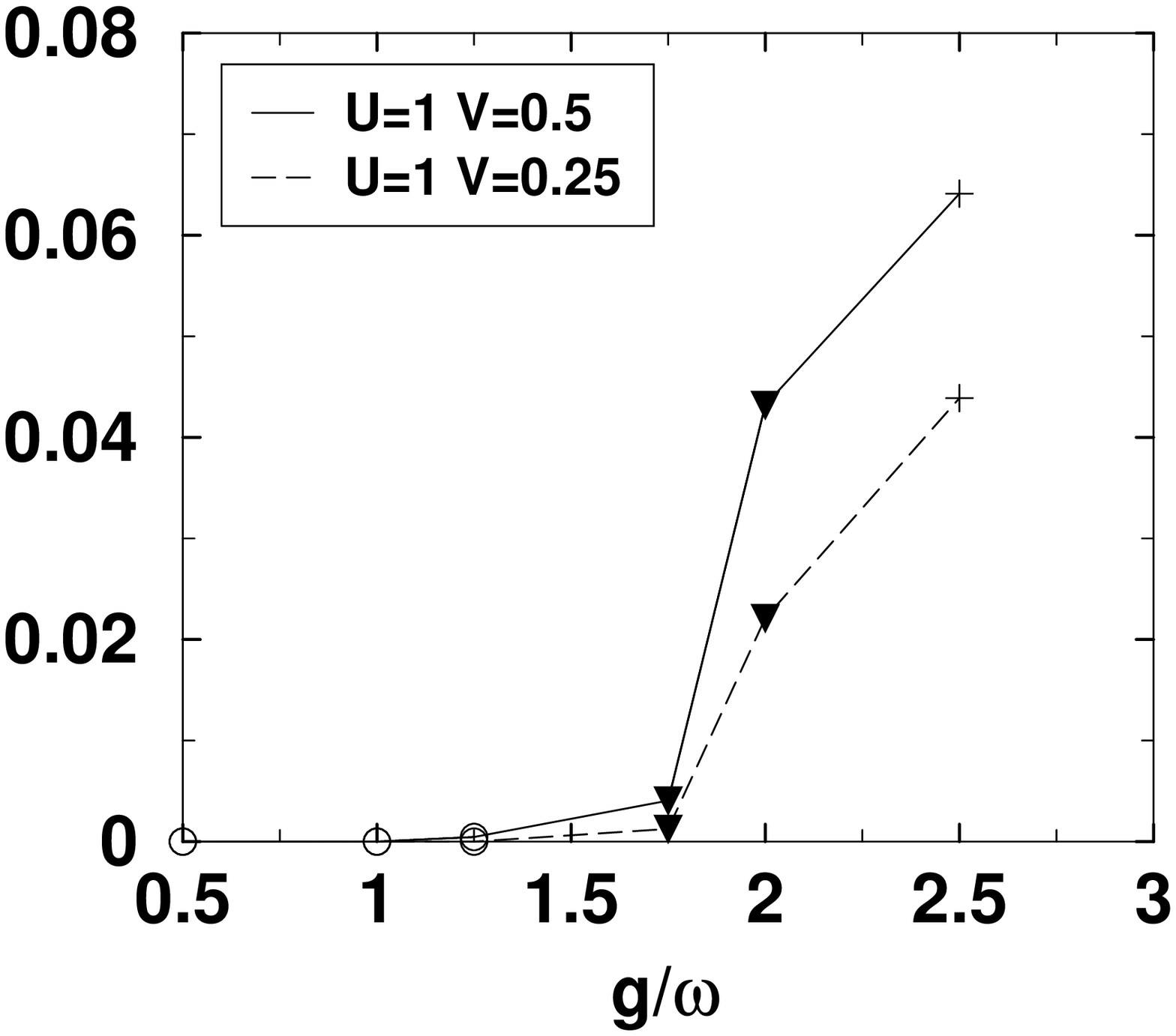}}}
\vspace{10mm}
\caption{Order parameter $X_{2k_F}$ as a function of $g/\omega$ for
$U=1$. The solid lines correspond to $U/V=2$ and the long dashed lines
to $U/V=4$. The symbols follow the same pattern as in the phases
diagrams.  }
\label{fig:x2kf}
\end{figure}

When $g/\omega$ increases toward the intermediate regime ($g/\omega >
1.5$), the system undergoes a MIT toward an insulating phase
presenting a $2k_{F}$ long range order. One should notice that this
phase is induced by the NN repulsions and does not exist in the
Hubbard Holstein model.  In comparison to the $V=0$ case, the $2k_F$
LRO phase develops at the expense of the bi-polaronic phase . It is
interesting to point out that for positive $U_{e\!f\!f}$ the
development of the $4k_F$ LRO phase, induced by the NN repulsions,
have a tendency to localize the electronic structure, while for
negative $U_{e\!f\!f}$, the development of the $4k_F$ LRO phase
corresponds to a tendency toward a less localization.  The amplitude
of the charge correlation functions ${\cal C}_{n}(j)$ extrapolate at
infinite inter-site distances toward finite values (for instance
$0.06$ for $U=1$, $V=0.5$ and $g/\omega=2$).  The order parameter has
been defined in the usual way
$$X_{2k_F} = \lim \limits_{N_{s}\rightarrow+\infty}
|\sum_{j}{e^{i2k_Fj}{\cal C}_{n}(j)}|$$ and is reported on
figure~\ref{fig:x2kf}. One can see that, as in the $4k_{F}$ LRO phase,
the order parameter increases very slowly at the MIT as in a
Kosterlitz-Thouless transition.

\begin{figure}[h]
\centerline{\resizebox{6.5cm}{6.5cm}
   {\includegraphics{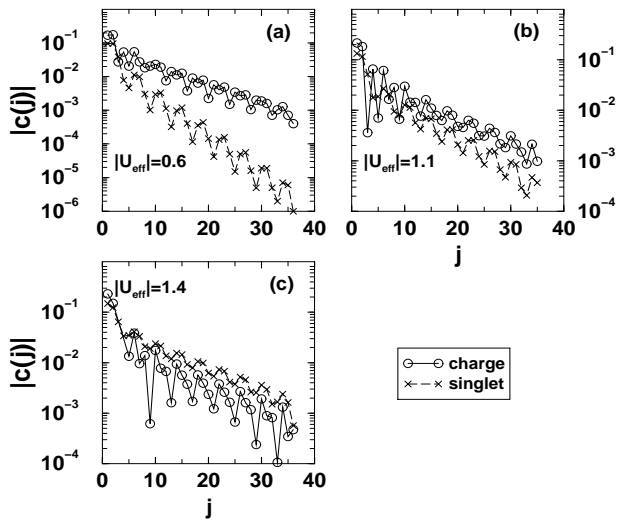}}}
\vspace{10mm}
\caption{Charge (circles) and singlet (crosses) fluctuations
correlation functions for $U/V=2$, $g/\omega=2$ and different values
of $U_{e\!f\!f}$. $U_{e\!f\!f}= 0.6$ (a), $U_{e\!f\!f}= 1.1$ (b) and
$U_{e\!f\!f}=1.4$ (c).}
\label{fig:ll1}
\end{figure}

Both spin and charge channels are gaped. It is noticeable that the
gaps values are always of the same order of magnitude~:
$\Delta_{\rho}=\Delta_{\sigma}=1.04$ for $U=0.2$, $V=0.1$ and
$g/\omega=2$, $\Delta_{\rho}= 0.35$, $\Delta_{\sigma}= 0.33$ for
$U=1$, $V=0.25$ and $g/\omega=2$.  Coherently, the fluctuations
correlation functions (spin, charge and on-site singlet) decrease
exponentially with the inter-site distances (see
figure~\ref{fig:ll1}).

One should note that, for large $|U_{e\!f\!f}|$, the on-site singlet
fluctuations correlations are dominant, whereas, for weak
$|U_{e\!f\!f}|$, the charge fluctuations correlations dominate.  On
the contrary the increase of $V$, increases the CDW at the expense of
the singlet fluctuations.

\subsubsection{bi-polaronic phase}
For large values of $g/\omega$ the system goes in a bi-polaronic phase
where the attractive nature of the effective on-site interaction
strongly couples the electrons in pairs.  This pairing is associated
with a strong localization and a self-trapping of the pairs, due to
the rescaling of the hopping integrals between low energy vibronic
states by Franck Condon factors. The remaining delocalization
processes are very small and the ground state wave functions have
dominant configurations of the type
$$..20002000..$$ where $2$ stands for a double occupancy of a site
and $0$ for an empty site. The probability of having a lonely electron
on a site is extremely small with for instance values smaller than
$10^{-7}$ for $U/t=1$, $V/t=0.25$ and $g/\omega=2.5$.  The energy per
site is nearly constant and in excellent agreement with the formula
$1/4\, U_{e\!f\!f} - 1/2\, g^2/\omega + 1/2\, \omega$ (the difference
between the computed values and the formula being of the oder of $
10^{-5}$ for $U/t=1$, $V/t=0.25$ and $g/\omega=2.5$).  Let us notice
that the GS is strongly quasi-degenerated due to the equivalence
between the four different phases of the CDW and to the absence of
second neighbors repulsion terms. As in the polaronic phase the
degeneracy splitting scales as $t_{e\!f\!f}$. Finally the system is
strongly gaped both in the charge and spin channels. Indeed, either to
extract an electron from the system or to build a triplet state, one
needs to break an electron pair. Such a mechanism cost the energy
$U_{e\!f\!f}$ and therefore both gaps scale as it.

\subsection{The basis set truncation effect}

\begin{figure}[h]
\centerline{\resizebox{6cm}{6cm}{\includegraphics{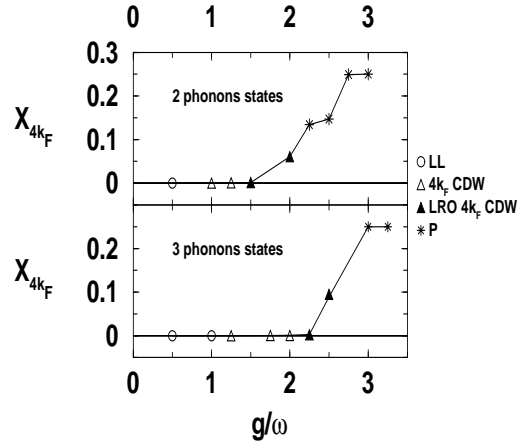}}}
\vspace{5mm}
\caption{$X_{4k_F}$ order parameter for two (upper curve) and three
(lower curve) phonons states kept per electronic state on each
site. The phases symbols follow the same pattern as in the phases
diagrams.}
\label{fig:23ph} \vspace*{2eM}
\end{figure}

As we already mentioned the phononic basis set truncation to two
vibronic states per occupation number should induce a bias toward
excessive localization in the intermediate regime. This is precisely
 the range of coupling parameters where the new phases appears and,
as we will see later,  the coupling regime which is the most
interesting for the physics of real systems. In order to quantify the
effects of the basis set truncation we have performed, in addition to
the small systems calculations presented in section~\ref{ss:cd},
infinite system DMRG calculations with three vibronic states kept for
each site occupation and spin. Each site is then described by twelve
states instead of eight. Since these calculations are very expansive
we have only performed them for a fixed set of electronic parameters
$U/t=4$ and $V/U=1/4$. Varying the electron-phonons coupling parameter
therefore leads us to go from the Luttinger Liquid phase up to the
polaronic phase. Figure~\ref{fig:23ph} reports the order parameter
$X_{4k_F}$ for the two and three phonons states.
As expected the increase of the basis set results in a displacement of
the phases transitions toward larger value of the coupling constant,
that is a larger delocalization of the system for a given value of
$g/\omega$. The Luttinger Liquid phase is only slightly enlarged, in
agreement with the weak participation of excited vibronic states in
the small systems wave functions. The phase which is the most favored
by the basis set increase is the metallic $4k_F$ CDW phase. Indeed
this phase is substantially extended toward larger values of
$g/\omega$ while the insulating LRO $4k_F$ phase is essentially
translated by $0.75g/\omega$.

In conclusion the basis set truncation does mot change the structure
of the phases diagram, its main effect being of reducing the range of
the metallic $4k_F$ CDW phase, essentially on the side of the larger
values of the electron-phonons coupling parameter, and shifting the
LRO $4k_F$ phase toward smaller values of $g/\omega$. All these results are in
complete agreement with the small system calculations.

\section{Relevance to the organic conductors physics}
In this section we are going to discuss the pertinence of the above
calculations for the low energy physics of organic conductors and more
specifically for the $\left(DX-DCNQI\right)_2M$ family and the
Bechgaard salts family. Indeed, as already mentioned in the
introduction, the building molecules of these systems present a
certain number of similar characteristics that are crucial for the
relevance of the molecular vibrations to the electronic structure. The
$DX-DCNQI$ as well as the $TMTTF$ or $TMTSF$ molecules are planar,
strongly conjugated and their skeleton is based on organic cycles~;
the quinone cycle for the $DX-DCNQI$ and the two pentagonal cycles of
the fulvalene for the $TMTTF$ and $TMTSF$. These geometrical
properties strongly favor the existence of low frequency vibrational
modes, namely the angular distortion of the cycles, the bond length
remaining untouched. Among these, the $A_g$ modes couple to the
electronic structure in a relatively strong manner, leading to e-mv
coupling constants in the intermediate range (as can be seen in Raman
experiments~\cite{vib} or in the geometry relaxation of the molecules
with their ionization~\cite{geomrlx}). In addition, one should notice
that due to their conjugated character, the $\pi$ system of the
considered molecules are strongly delocalized on their skeleton and
thus strongly polarizable. The consequence is that the second neighbor
coulombic interactions should be strongly screened by the ``metallic
plate'' of the in-between molecule. One can therefore reasonably
assume that the on-site and first neighbor repulsion terms are the
only pertinent ones for the physics of these organic 1D systems.

Recently charge ordered phases presenting similar characteristics with
the $4k_F$ LRO CDW state have been observed both in the
$\left(DX-DCNQI\right)_2M$ family and in the Bechgaard salts family.

The most characteristic compound is certainly the
$\left(DI-DCNQI\right)_2A\!g$ compound. Indeed, this system which is
strongly one dimensional undergoes a MIT phases transition at $220K$
toward a $4k_F$ CDW state~\cite{dcnqi2}. Both NMR~\cite{dqirmn} and
X-ray~\cite{dqirx} show that the insulating state exhibit a on-site
charge disproportionation between two adjacent molecules. This charge
order saturates at a $3:1$ ratio below $140K$ and is associated with a
molecular geometry deformation due to the ionicity modification.  In
addition, the spin channel remains ungaped~\cite{dqirmn}. This
compound, which is usually considered as strictly non-dimerized
(despite a recent doubt raised by Meneghetti {\it et
al}~\cite{dcnqidim}) seems to be the perfect example of an e-mv driven
$4k_F$ LRO CDW state as the one discussed in this paper.  Most authors
assume that this LRO state is driven by a Wigner crystallization due
to strong long range electronic repulsions. We have seen in the
previous sections, that the e-mv driven $4k_F$ CDW phases present very
similar characteristics as the Wigner crystals, without the need of
very strong correlation effects and without long range repulsions.  In
the light of the previous considerations on the strongly polarizable
electronic structure of the quinone cycles, on the usually assumed
 values of the electronic correlation strength (in the
intermediate range rather than the strong range for the $DCNQI$
family~\cite{DCNQIdft}), and on the e-mv coupling characteristics, it
seems to us more plausible that the considered $4k_F$ CDW state is
driven by the e-mv than by the usual unscreened strong coulombic
repulsions.  At very low temperatures the
$\left(DI-DCNQI\right)_2A\!g$ undergoes a second phases transition
toward a spin ordered state where the $4k_F$ LRO CDW is associated
with a $2k_F$ anti-ferromagnetic LRO. These results are in agreement
with the $2k_F$ SDW fluctuations exhibited in the $4k_F$ LRO CDW phase
of the eHH model, fluctuations that could easily be pinned at low
temperatures by impurities or inter-chains interactions.

Another family for which the preceding analyses are relevant about the
e-mv influence on the electronic structure, is the Bechgaard salts
family.  This fact should be put in the light of the newly discovered
$4k_F$ charge ordered phase. This phase have been observed on several
systems such as in the $\left(TMTTF\right)_2 PF_6$ or the
$\left(TMTTF\right)_2A\!sF_6$ ~\cite{nmrpf6}. The charge ordered (CO)
phase appears when the temperature is lowered from the metallic
Luttinger Liquid phase through a soft cross over.  It is noticeable
that transport measurements observe $4k_F$ CO fluctuations in the
normal LL phase, indicating that the interactions driving the
transition are relevant far inside the normal phase and over a large
range of pressures~\cite{transp,diel1}.  X-ray diffraction
measurements do not see any n-merization associated with this phases
transition, excluding a Peierls mechanism. NMR measurements on the
carbon atoms of the central double bond of the fulvalene (on which the
Highest Occupied Molecular Orbital, responsible for the low energy
physics of these systems, have large coefficients) exhibit a charge
disproportionation on NN molecules. Magnetic susceptibility
measurements are transparent to this phases transition and the spin
channel remain ungaped. It is clear that all these experimental
results are in full agreement with the characteristics found in the
present work for the $4k_F$ LRO phases transition. Despite the fact
that the system dimerization has not been taken into account in the
present work, the e-mv coupling appears as a strong candidate for the
CO driving mechanism. It is however clear that the dimerization degree
of freedom should be taken into account, in addition to the
correlation degrees of freedom (both on-site and inter-sites) and the
e-mv degrees of freedom treated in the present work, for a complete
description of the Bechgaard salts.

Finally, one can point out that the mechanism underlying a MIT
transition toward an on-site charge modulation, driven by the e-mv,
does not imply any lattice distortion apart from small modifications
in the molecular geometries, in particular in the angles of the cycles
(the fulvalene cycles in the Bechgaard salts). Indeed such geometrical
relaxations can be expected to be a consequence of the charge
disproportionation. These transitions should therefore be $q=0$ transitions.

\section{conclusion}
The present paper studies the influence of the electron-molecular
vibration coupling on the electronic structure of correlated 1D
quarter-filled chains. The model chosen is the extended-Hubbard
Holstein model, that is the simplest one containing both electron
correlation effects relevant for the physics of molecular crystals
such the organic conductors, and e-mv coupling whose effects can be
expected to play a crucial role. We have found that for low phonons
frequencies the electronic structure is strongly affected by the
presence of molecular vibrations. One of the most striking results
being the existence of $4k_F$ CDW phases, one metallic and one
insulating charge ordered phase, for small values of the correlation
strength (as small as $U/t=2$), small values of the nearest neighbor
repulsion and in the absence of long range coulombic repulsion.  A
study of these phases shows however that they have similar
characteristics as a Wigner crystal. In this light a new
interpretation ---~based on the e-mv coupling~--- of the origin of the
$4k_F$ charge ordered phase in the $(DI-DCNQI)_2A\!g$ and of the
recently discovered CO phase in the $(TMTTF)_2X$ family has been
proposed. Such an interpretation of the apparition of the $4k_F$ CO
phase present the advantage over the usual purely electronic
interpretation to be in agreement with the usual values of the
correlation strength for these systems and the strong screening of the
long range bi-electronic repulsions that can be expected from the
$\pi$ conjugated character of the building molecules. To conclude we
would like to point out that for a complete description of the
Bechgaard's salts physics, one should treat in addition to the
intra-molecular phonons modes, the inter-molecular modes responsible
for the known dimerization in these compounds. In view of the recent
work of Campbell, Clay and Mazumdar~\cite{CCM} on the adiabatic
dimerized extended-Hubbard Holstein model that forecasts the existence
of mixed phases such as the Bond Charge Density Wave and the
Spin-Peierls $4k_F$ CDW phase, it would be of interest to conduct the
same type of study as the present one on the dimerized extended-Hubbard
Holstein model. Indeed, such a model would include all degrees of
freedom important for 1D organic conductors as well as the phonons
quantum fluctuations that have been proved to be crucial for a correct
description of the e-mv coupling~\cite{sol}

\end{document}